\documentclass[twocolumn,showpacs,showkeys,preprintnumbers,amsmath,amssymb]{revtex4}

\usepackage[dvips]{graphicx}
\usepackage{dcolumn}
\usepackage{bm}

\begin{document}

\preprint{}

\title{Observation of modified radiative properties of cold atoms \\
in vacuum near a dielectric surface}

\author{V.~V. Ivanov, R.~A. Cornelussen, H.~B. van Linden van den Heuvell, 
and R.~J.~C. Spreeuw}
\email{spreeuw@science.uva.nl}
\affiliation{Van der Waals-Zeeman Institute, University of Amsterdam, \\
Valckenierstraat 65, 1018 XE Amsterdam, The Netherlands}
\homepage{http://www.science.uva.nl/research/aplp/}

\date{\today}

\begin{abstract}

We have observed a distance-dependent absorption linewidth of cold
$^{87}$Rb atoms close to a dielectric-vacuum interface. This is the
first observation of modified radiative properties in vacuum near a
dielectric surface. A cloud of cold atoms was created using a
magneto-optical trap (MOT) and optical molasses cooling. Evanescent
waves (EW) were used to observe the behavior of the atoms near the
surface. We observed an increase of the absorption linewidth with up to
25\% with respect to the free-space value. Approximately half the
broadening can be explained by cavity-quantum electrodynamics (CQED) as
an increase of the natural linewidth and inhomogeneous broadening. The
remainder we attribute to local Stark shifts near the surface. By
varying the characteristic EW length we have observed a distance
dependence characteristic for CQED.

\end{abstract}

\pacs{42.50.-p, 42.50.Xa, 42.50.Pq}

\keywords{Vacuum field fluctuations, cold atoms, spontaneous emission}

\maketitle

\section{Introduction}

An electronically excited atom (or molecule) can decay to the ground
state by spontaneous emission. The characteristic rate at which this
occurs is not simply an intrinsic property of the atom but also depends
on the environment. The spontaneous emission rate is proportional to the
density of electromagnetic field modes (DOS, or ``density of states''),
which is determined by the electromagnetic boundary conditions. The DOS
can thus be modified, and with it the spontaneous emission rate. The
boundary conditions imposed by the environment not only change the
radiative linewidth but also induce energy level shifts and thus change
the transition frequencies. These include the electrostatic or Van der
Waals shift, the Casimir-Polder shift (modification of the Lamb shift),
and resonant radiative shifts. For a review see, e.g. Ref.~\cite{Hin94}.

Modified spontaneous emission was first observed by Drexhage
\cite{Dre70,Dre74}, using dye monolayers separated from an interface by
fatty acid layers. Both inhibited and enhanced spontaneous emission have
since then been observed by others in a variety of geometries and
circumstances \cite{HulHilKle85,HeiChiTho87,VreHunSch93,SnoLagPol95}.
Remarkably, the radiative linewidth of an atom {\em in vacuo} at a
distance of the order of an optical wavelength from a dielectric surface
has never been investigated experimentally. Energy level shifts have
been studied for atom inside cavities \cite{SanSukHin92,SukBosCho93} and
in vapor cells, using selective reflection spectroscopy
\cite{CheBloRah91,FaiSalFic03}. The situation of an atom in front of a
distant mirror has recently been investigated using a single trapped
ion. Both the broadening of the radiative linewidth and energy level
shifts have been reported for this system
\cite{EscRaaSch01,WilBusEsc03}.

In this paper we experimentally investigate the radiative properties of
cold ($T\approx 10\,\mu$K) atoms of $^{87}$Rb close to a glass surface,
at a distance on the order of an optical wavelength. Using our method of
evanescent-wave spectroscopy \cite{CorAmeWol02} we have observed a
broadening of the absorption linewidth. We compare our observations with
calculations based on CQED.

\section{Method: evanescent-wave spectroscopy}

The radiative linewidth $\Gamma$ is proportional to the power spectral
density of the vacuum field fluctuations at the position of the atom
\cite{KhoLou91}, i.e. the local DOS. The proximity of a dielectric
surface imposes a boundary condition on the field, changing the DOS.
This leads to a modification of $\Gamma$ and to energy level shifts
\cite{KhoLou91,CheBloRah91,HinSan91,SnoLagPol95,CouCouMer96,WuEbe99,
FaiSalFic03}. Both the linewidth broadening and the level shifts are
significant mainly at distances $z\lesssim\lambdabar=\lambda/2\pi$,
where $\lambda$ is the wavelength of the dominant electronic transition.
In our case this is the $D_2$ resonance line of Rb, and
$\lambdabar=124\,$nm.

Therefore we probe the cold atoms near the glass surface using
evanescent-wave (EW) spectroscopy \cite{CorAmeWol02}. This method is
selectively sensitive to atoms very close to the surface. An EW appears
when our probe beam undergoes total internal reflection at the glass
surface with index of refraction $n=1.51$, see Fig.~\ref{fig:setup}. The
optical field on the vacuum side decays exponentially with the distance
$z$ to the surface, $E(z)\propto\exp(-z/\xi)$. Atoms can absorb light
from the EW, if their distance to the surface is on the order of the
decay length $z\lesssim\xi\sim\lambda$. The decay length can be adjusted
by changing the angle of incidence $\theta$ according to
$\xi(\theta)=\lambdabar(n^2\sin^2\theta-1)^{-1/2}$. By adjusting
$\theta$ we can thus vary the distance scale at which the atoms interact
with the probe light. By increasing $\theta$ further above the critical
angle $\theta_c=\arcsin n^{-1}$, the absorption will occur closer to the
surface, where $\Gamma$ is more strongly modified.

For atoms in free space, the absorption profile is given by a Lorentzian
profile, centered at the (free-space) atomic transition frequency
$\omega_{eg}=c/\lambdabar$ and with a full width at half maximum (FWHM)
equal to the natural linewidth $\Gamma_{\infty}/2\pi=6.07\,$MHz. Both
$\omega_{eg}$ and $\Gamma_{\infty}$ change in the proximity of the
surface. Roughly speaking these $z$-dependent Lorentzians become
convoluted with the EW energy density $U(z)\propto\exp(-2z/\xi)$.
Therefore, in our experiment $\omega_{eg}(z)$ also contributes to the
observed absorption linewidth through inhomogeneous broadening. We
expect the width of the resulting absorption profile to increase with
the angle of incidence $\theta$. In the experiment we measured this by
tuning an EW probe laser across the profile and measuring the
absorption.

\section{Experiment}

\begin{figure}[ht]
\includegraphics[width=50mm]{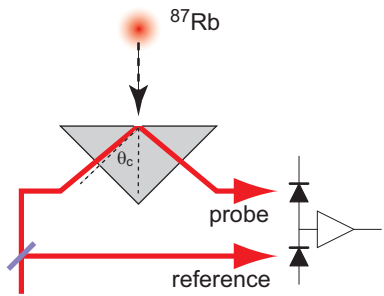}
\caption{\label{fig:setup} Scheme of the experiment. A weak
evanescent-wave probe beam is reflected from the prism surface and
collected on a photodiode. The reference beam has equal power but
bypasses the vacuum cell. The difference photocurrent of probe and
reference beams yields the absorption signal, typically a fraction of
$10^{-3}-10^{-4}$ of the probe. A second reference beam (not shown) was
used to monitor variations in the probe power. We also collect the
fluorescence from the MOT to normalize for shot to shot variations of
the number of atoms.}
\end{figure}

The major part of our experimental setup has been described previously
\cite{VoiWolCor01}. We produced clouds of cold $^{87}$Rb atoms using
magneto-optical trapping inside a ultra high vacuum cell (base pressure
$p\simeq{10}^{-10}$mbar). After postcooling in optical molasses we ended
up with about $3\times 10^7$ atoms, at a temperature of 9~$\mu$K. At
this temperature the Doppler width is 90~kHz (FWHM). The cooling lasers
were then switched off and the atoms fell down toward the surface of a
glass prism, about 3.6~mm below. The center of the cloud reached the
prism surface and the EW spot after 27~ms. Just before hitting the
surface, the atoms briefly interact with a weak, $p$-polarized, EW probe
beam, see Fig.~\ref{fig:setup}. The intensity of the probe was kept well
below the saturation intensity to avoid power broadening. Using
$0.35\,\mu$W and a waist of about 1~mm, the maximum saturation parameter
was $s\simeq 0.08$.

The probe beam was derived from a home-built diode laser system, locked
to the $F=2\rightarrow F'=(1,3)$ crossover resonance in the $D_2$ line
of $^{87}$Rb (780 nm). We used an acousto-optic modulator (AOM) to shift
the probe frequency near resonance with the $F=2\rightarrow F'=3$
transition and to tune it across the resonance. Before sending it into
the cell, a fraction of the probe beam was split off and sent to a
photodiode as a reference. After total internal reflection on the prism
surface the probe beam was focused on a second photodiode. The
photocurrents of the two photodiodes were subtracted to obtain our
signal, typically a fraction of $10^{-3}-10^{-4}$ of the probe.

The difference photocurrent was amplified by a low-noise current
amplifier (Femto, LCA-100K-50M, 50~MV/A transimpedance) and sent through
a low-pass filter ($RC=1\,$ms) to further reduce the noise. All
photodiode signals, including a power monitor and a MOT fluorescence
monitor were acquired using a digital storage oscilloscope. The latter
two signals were used to normalize the absorption signals for variations
in the probe power and shot-to-shot variations in the number of cold
atoms.

\begin{figure}
\includegraphics[width=70mm]{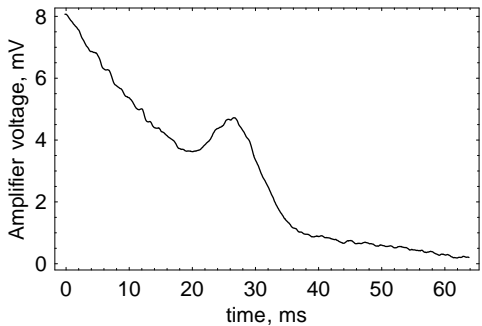}
\caption{\label{fig:time-trace}
A typical time-of-flight signal taken
with the evanescent-wave probe beam (100$\times$ average). The peak
around 27~ms is due to the absorption of evanescent probe light by cold
atoms arriving at the surface. As a result of high pass filtering the
signal has been distorted and sits on top of an exponentially decaying
transient ($1/e$ time 26~ms).}
\end{figure}

In Fig.~\ref{fig:time-trace} we show a typical EW absorption time trace
with the probe beam tuned near resonance, for an angle of incidence
$\theta_i=\theta_c+0.52^\circ$. The signal has been averaged 100$\times$
to reduce the noise. Without filtering, the absorption signal has a
Gaussian shape due to the velocity distribution of the falling atoms.
However, in order to suppress slow drifts in the difference photocurrent
we used AC coupling (i.e. a high-pass filter) on the oscilloscope. As a
result the Gaussian signal has been distorted. Furthermore it is
superposed on an exponentially decaying transient originating from
switching off the MOT/molasses beams. Although we shielded the
photodiodes from the molasses light as much as possible, some light is
still detected. Unfortunately the time between switching off the lasers
and the arrival of the atoms at the surface is fixed by gravity. 

The position of the peak corresponds to the fall time of the atoms. The
width is given by the ratio of the size ($\sim 3$~mm) and velocity
($\sim 0.3$~m/s) of the atom cloud as it reaches the surface. The height
of the peak is $\sim 2$~mV, which corresponds to an absorbed power of
$\sim 80$~pW. The time-integrated signal amounts to $\sim 3\times10^6$
absorbed photons, or $\sim 2$ scattered photons per atom in the center
of the EW. If we tune the probe laser away from resonance, or increase
the angle of incidence, the signal amplitude decreases and the number of
scattered photons drops to much less than one per atom. Eventually the
signal disappears in the noise, which is dominated by shot noise.

\section{Data processing}

Despite the signal distortion, we can extract the amplitude and width of
the original Gaussian, by fitting the filtered time trace to an
analytical expression. This expression involves error functions due to
the known step response function of the filter. We took the fitted
height of the Gaussian as the measure for the amount of absorption. The
Gaussian width is essentially constant. Thus, the height of the peak is
proportional to the absorbed EW power, which depends on the EW detuning
and the angle of incidence.

\begin{figure}
\includegraphics[width=80mm]{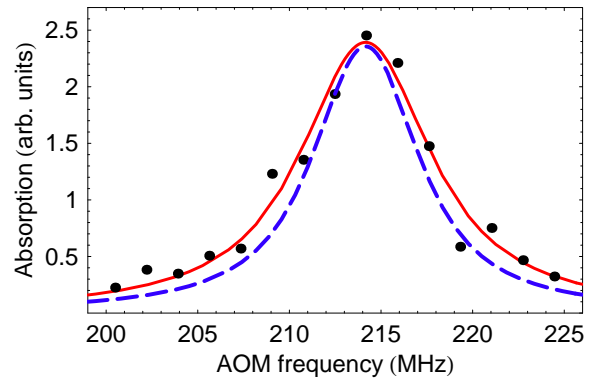}
\caption{\label{fig:absprofile} Measured and fitted absorption profiles.
The AOM frequency is the shift imparted to the frequency of the locked
probe laser to tune it near resonance. Each data point is the fitted
Gaussian amplitude of a 100$\times$ averaged time trace as in
Fig.~\ref{fig:time-trace}. The solid line is the fitted Voigt profile.
For comparison, the dashed line shows the free-space Lorentzian
profile.}
\end{figure}

For a given angle of incidence we measured time traces for different
detunings of the EW probe. The fitted Gaussian height as a function of
probe detuning yields an absorption profile as shown in
Fig.~\ref{fig:absprofile}. From this we extracted a Lorentzian line
width by fitting a Voigt profile
\begin{multline}
\label{eq:voigt}
\mathcal{V}(\omega)=
\frac{A}{\sqrt{2\pi}\Delta}\mbox{Re}
\left\{
\exp \left[ -
\left( \frac{\omega-\omega_{eg}+i\Gamma/2}{\sqrt{2}\Delta}\right) ^2\right]
\times\right. \\
\left. \times\mbox{erfc}
\left( -i\frac{\omega-\omega_{eg}+i\Gamma/2}{\sqrt{2}\Delta}
\right)
\right\},
\end{multline}
where $A$ is an amplitude and $\mbox{erfc}(.)$ is the complementary
error function. This Voigt profile is the convolution of a Gaussian with
a fixed width $\Delta/2\pi=1\,$MHz and
a Lorentzian with variable width $\Gamma$.

The fixed Gaussian linewidth accounts for the finite spectral width of
the probe laser. Our grating stabilized diode laser system has a
spectral linewidth comparable to the observed atomic linewidth
broadenings \cite{TurWebHaw02,WieHol91}. We determined the laser
linewidth in a separate experiment by observing the beat note between
two similar but independently locked diode lasers on a photodiode. The
observed decoherence of the beat signal was well described by a Gaussian
with a width of 1~MHz. This is the linewidth at short (ms) time scales.
For longer timescales we rely on the feedback loop of the laser locking
electronics.

\begin{figure}
\includegraphics[width=85mm]{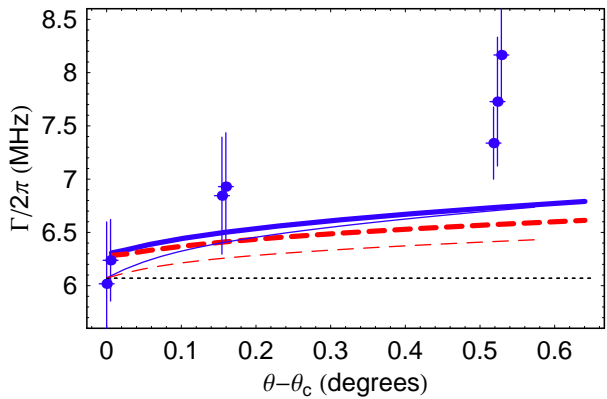}
\caption{\label{fig:finalplot} (Color online) Fitted linewidths
for varying angle of incidence of the evanescent-wave (EW) probe.
The data points at $\theta-\theta_{c}=0$ have been measured in
free space, instead of with an EW probe. The thin lines are the
calculated widths based on Eq.~(\ref{eq:abs}), the thick lines
based on an integration of the optical Bloch equations (see text).
The dashed curves (red) show the result when level shifts are not
taken into account, the solid curves (blue) take into account both
broadening and level shifts.}
\end{figure}

The Lorentzian linewidth $\Gamma$ contained in the Voigt profile is a
fit parameter. We performed measurements of the absorption profile
several times for two different angles of incidence
($\theta-\theta_{c}=0.16^{\circ}$ and $0.52^{\circ}$). For each
absorption profile we find one value for $\Gamma$. The results are shown
in Fig.~\ref{fig:finalplot}. For larger angles, the absorbed power
became too small compared to the noise, due to the decreasing EW volume.
The vertical error bars in Fig.~\ref{fig:finalplot} are entirely
determined by the scatter of the datapoints as seen in
Fig.~\ref{fig:absprofile}.

In the limit of large EW decay length, or $\theta\rightarrow\theta_{c}$,
we expect $\Gamma$ to tend to the free-space value $\Gamma_{\infty}$.
Unfortunately, at angles very close to the critical angle,
$\theta-\theta_{c}<0.05^{\circ}$, the probe beam cannot be treated as a
plane wave due to the finite diffraction angle. In order to avoid this
complicated situation, we performed an independent check by measuring
the linewidth in free space. The same probe laser was used to measure
the absorption by the atomic cloud while falling, at a height of 2 mm
above the surface. A short flash of probe light was used to illuminate
the atoms. The probe beam containing a "shadow" due to absorption by the
atoms was recorded on a CCD camera. In addition to this picture (a), we
took two more: a background picture with no probe light (b), and a
reference picture with probe light but no atoms (c). The optical density
was then obtained as $\ln[(a-b)/(c-b)]$ and was measured for different
values of the probe detuning. The detuning was varied in a similar way
as in the EW probe measurements. The linewidth was again determined by
fitting a Voigt profile. We plotted this free-space value in
Fig.~\ref{fig:finalplot} as the datapoint for $\theta-\theta_{c}=0$. Our
measured free-space linewidth is in good agreement with the known value.
This shows that there were no unknown systematic broadening effects,
such as saturation, stray magnetic fields, or residual Doppler
broadening.

\section{Theory for $\Gamma(z)$}

We will now compare the measured $\theta$-dependence of the
linewidth to CQED calculations. When the atom approaches the
dielectric surface, both the radiative linewidth $\Gamma$ and the
resonance frequency $\omega_{eg}$ change in a $z$-dependent way.
Although the latter does not change the excited state lifetime, 
it appears as inhomogeneous broadening in the experiment,
because the evanescent wave performs an integration over $z$.

The absorption from the EW probe beam can be calculated by performing a
spatial integration of the photon scattering rate over the vacuum half
space $z>0$. The photons that are scattered out of the EW by atoms are
missing from the reflected probe beam so that the reflectivity drops
below unity. This approach works well if the absorption is small ($\ll
1$, i.e. no probe depletion). In the limit of low saturation the photon
scattering rate is $\Gamma(z)s(z)/2$, with the saturation parameter
given by
\begin{equation}
s(z)=\frac{c\, U(z)}{I_0}\;\frac{\Gamma^2_\infty/4}{\delta^2(z)+\Gamma^2(z)/4}\,,
\end{equation}
where $U(z)\propto\exp(-2z/\xi)$ is the EW energy density and
$I_0=1.6\,$mW/cm$^2$ is the (free space) saturation intensity. Note that
an increase of $\Gamma(z)$ not only increases the Lorentzian width but
also multiplies into the photon scattering rate, thus increasing the
on-resonance rate. This effect tends to favor the detection of atoms
near the surface.

The modification of the radiative linewidth of an atom near a plane
dielectric surface has been described theoretically in terms of dipole
damping rates $\Gamma_{\perp}$ and $\Gamma_{\parallel}$
\cite{KhoLou91,CouCouMer96}. The subscripts $(\perp,\parallel)$ refer to
dipoles oriented perpendicular and parallel to the surface,
respectively. The dipole damping rates vary with the distance to the
surface $z$, in the notation of Ref.~\cite{CouCouMer96} (note also
\footnote{Note that there appears a printing error in Eq.~(58) of
\cite{CouCouMer96}; the coefficients $r^p(u)$ and $r^s(u)$ have been
interchanged.}):
\begin{eqnarray}
\frac{\Gamma_{\perp}(z)}{\Gamma_{\infty}} & = & 1+\frac{3}{2}\mathrm{Re}\int_0^{\infty}
\frac{u^3 r^p(u)du}{\sqrt{1-u^2}} \exp(2ikz\sqrt{1-u^2}),\nonumber\\
& & \label{eq:gperp}\\
\frac{\Gamma_{\parallel}(z)}{\Gamma_{\infty}} & = & 1+\frac{3}{4}\mathrm{Re}\int_0^{\infty}
\frac{u\,du}{\sqrt{1-u^2}} \left(r^s(u)+(u^2-1)r^p(u)\right)\nonumber \\
& & \times\,\exp(2ikz\sqrt{1-u^2}).
\label{eq:gpar}
\end{eqnarray}
Here $k=2\pi/\lambda$, and $r^{p}(u)$ and $r^{s}(u)$ are the Fresnel
reflection coefficients for $p$ and $s$ polarization:
\begin{eqnarray}
r^p(u) & = & \frac{n^2\sqrt{1-u^2}-\sqrt{n^2-u^2}}{n^2\sqrt{1-u^2}+\sqrt{n^2-u^2}},
\label{eq:rp}\\
& & \nonumber\\
r^{s}(u) & = & \frac{\sqrt{1-u^2}-\sqrt{n^2-u^2}}{\sqrt{1-u^2}+\sqrt{n^2-u^2}}.
\label{eq:rs}
\end{eqnarray}

\begin{figure}
\begin{flushright}
\includegraphics[width=72mm]{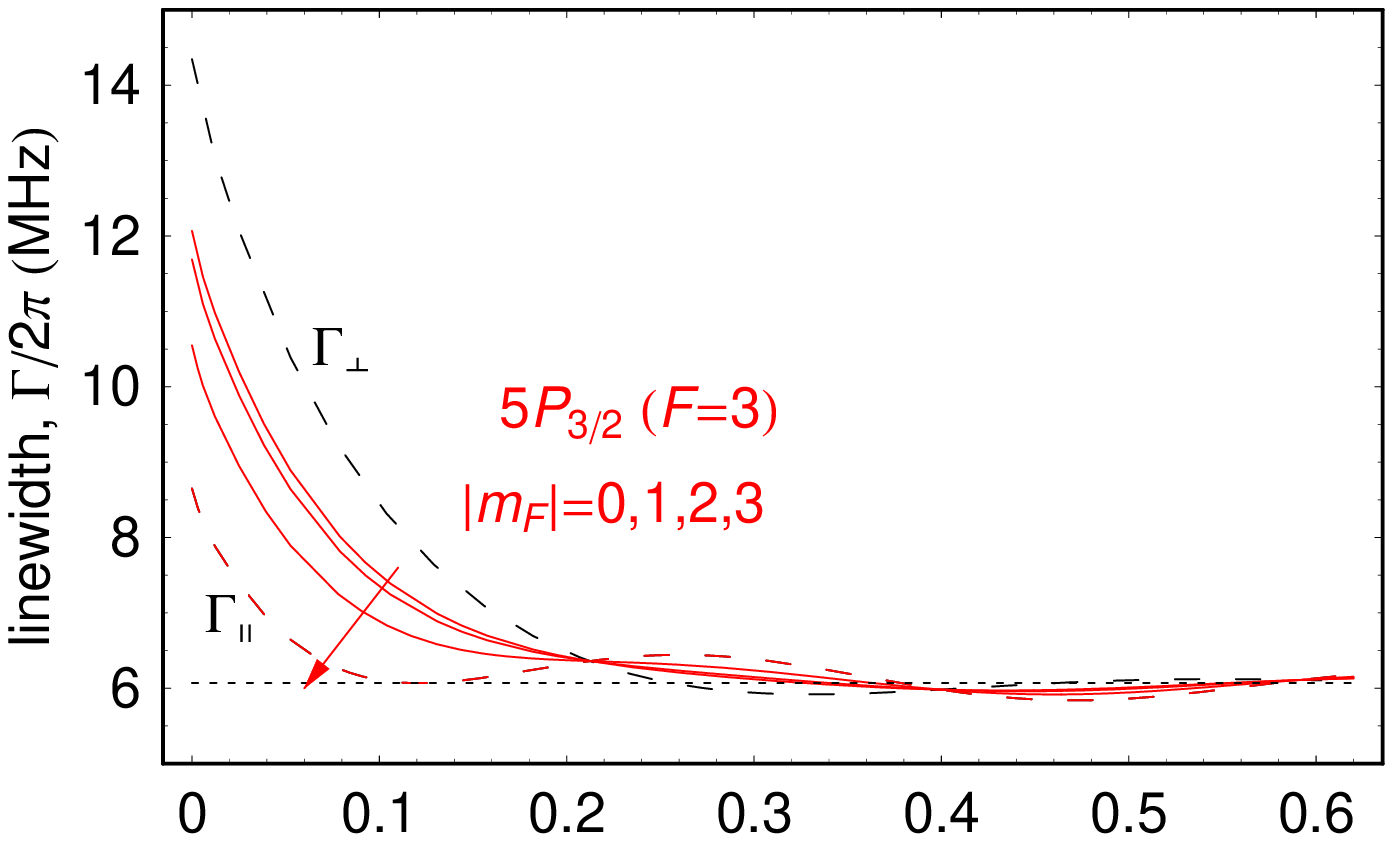}\rule{5mm}{0pt}\\
\includegraphics[width=80mm]{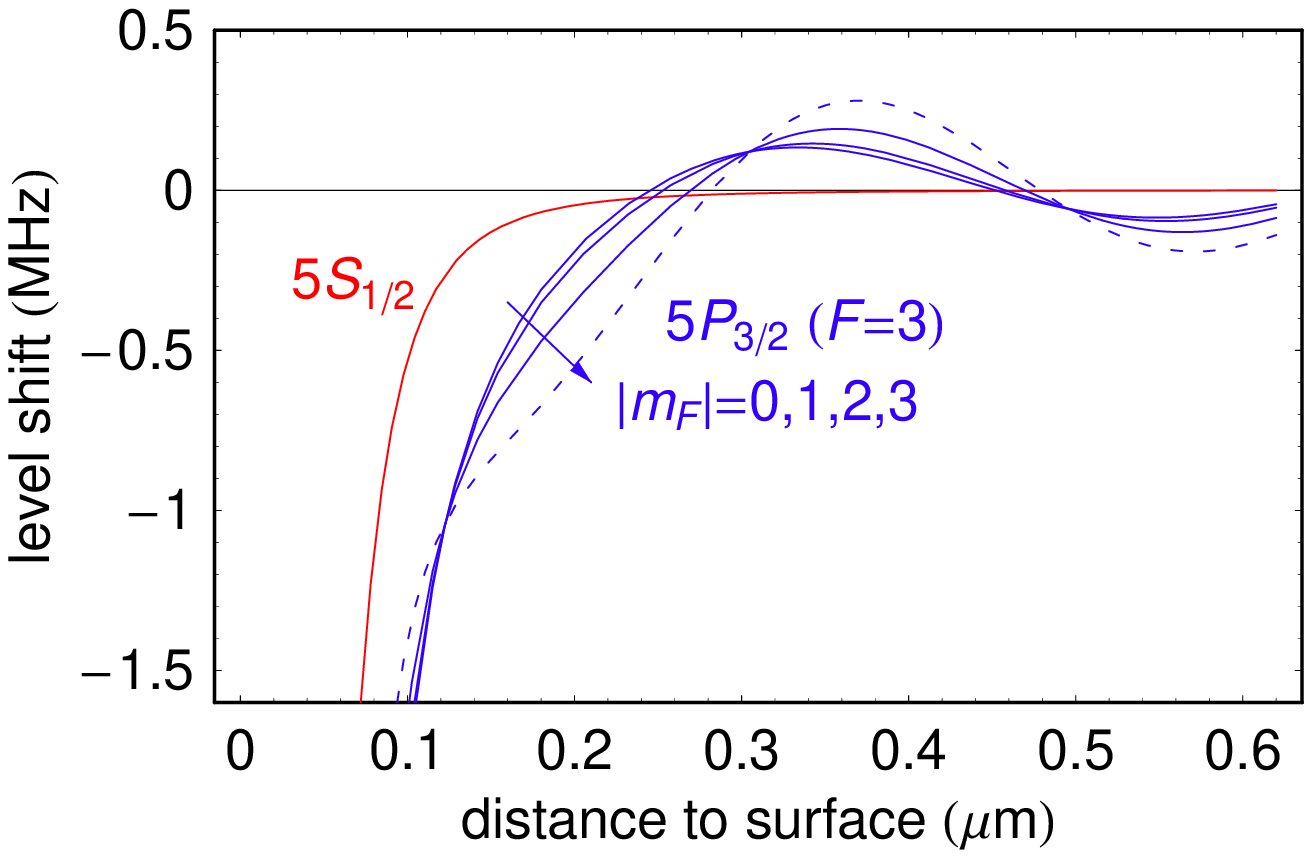}\rule{5mm}{0pt}
\end{flushright}
\caption{\label{fig:Spontaneous} Distance dependence of line widths
(upper) and level shifts (lower) of the relevant hyperfine magnetic
sublevels. In the upper graph, the dashed curves show the dipole damping
rates, see Eqs.~(\ref{eq:gperp}, \ref{eq:gpar}). The curves for the
$F=|m_F|=3$ states coincide with that for $\Gamma_\parallel$ and
do not contribute in the experiment. The dotted line is the free-space
value. In the lower graph the level shifts of the $F=|m_F|=3$
states do not contribute and are again shown as dashed.}
\end{figure}

In our experiment we probe $^{87}$Rb atoms on the transition
$5S_{1/2}(F=2)\rightarrow 5P_{3/2}(F'=3)$. An atom in the excited
magnetic hyperfine state $|F',m_F\rangle=|3,m\rangle$ can decay to the
ground state $|2,m-q\rangle$ with $q=0,\pm 1$. Choosing the quantization
axis perpendicular to the surface, the $q=0$ decay channel is governed
by $\Gamma_{\perp}$, the $q=\pm 1$ channels by $\Gamma_{\parallel}$. The
decay rate for a given sublevel $|3,m\rangle$ is then given by
\begin{equation}
\Gamma_m(z)=c_{m,0}\Gamma_{\perp}(z)+(c_{m,-1}+c_{m,1})\Gamma_{\parallel}(z),
\end{equation}
where $c_{m,q}$ is shorthand for the square of a Clebsch-Gordan
coefficient, $c_{m,q}=\langle 2,m-q,1,q|3,m\rangle^2$. Note that this
implies that close to the surface the $m$ states have different
lifetimes \cite{CouCouMer96}. The different $\Gamma_m(z)$ curves are
shown in Fig.~\ref{fig:Spontaneous}, together with $\Gamma_{\perp}(z)$
and $\Gamma_{\parallel}(z)$. The curve for $|m|=3$ is not relevant in
our experiment because our $p$-polarized probe does not excite these
$m$-states. Our $^{87}$Rb atoms are in a random mixture of all five
$|2,m\rangle$ states. The probe light is linearly polarized,
perpendicular to the surface, thus exciting $q=0$ transitions.

\section{Analysis; comparison with theory}

Integrating the photon scattering rate $\Gamma s/2$ over all $z$ and
averaging over the $m$-states, we arrive at the absorption of the probe;
expressed as a fraction:
\begin{equation}
-\frac{\Delta P}{P}\propto\int_0^\infty\sum_{m=-2}^{2}
\frac{\Gamma_m(z)\,c_{m,0}\,\rho(z)\,e^{-2z/\xi}}{\delta_m^2(z)+\Gamma_m^2(z)/4}\;dz,
\label{eq:abs}
\end{equation}
where $\rho(z)$ is the atom density. It is $z$-dependent due to the
ground-state level shift that accelerates the atoms to the surface. This
is well approximated by the Van der Waals potential $-C_3/z^3$,
resulting in a depletion of the density near the surface according to
$\rho(z)=\rho_0 (1+(z_W/z)^3)^{-1/2}$ with
$z_W=(C_3^g/m\,g\,h)^{1/3}\approx 50\,$nm. Here $C_3^g=5.6\times
10^{-49}\,$J~m$^3$ is the Van der Waals coefficient and $m\,g\,h$ is the
kinetic energy with which the atoms fall onto the surface.

The $z$-dependence of the laser detuning $\delta_m(z)$ in
Eq.~(\ref{eq:abs}) accounts for the energy level shifts by unequal
amounts for the ground and excited states. The shift of the ground
$5S_{1/2}$ state is dominated by the Van der Waals shift $-C_3^g/z^3$.
The shift of the excited $5P_{3/2}$ state is more complicated,
containing also a resonant component with oscillatory $z$-dependence. We
have used expressions for the shifts of both the ground and the excited
states from Ref.~\cite{HinSan91}, using transition line strengths taken
from Ref.~\cite{SafWilCla04}. We have extended the expressions from
\cite{HinSan91} to account for hyperfine structure. Furthermore we have
multiplied the results by a factor $(n^2-1)/(n^2+1)$, because our
surface is a dielectric instead of a mirror. This is known to be correct
in the nonretarded limit \cite{EbeWu03}, which gives the dominant
contribution in the experiment.

It is evident from Eq.~(\ref{eq:abs}) that the absorption profile is a
convolution of Lorentzians with different widths, amplitudes, and
central frequencies. The resulting absorption profiles are strictly
speaking no longer Lorentzian. We have numerically calculated the
expected absorption profiles using Eq.~(\ref{eq:abs}). In practice the
deviation from a Lorentzian is sufficiently small that we can fit a
Lorentzian to the calculated profiles. The fitted widths are plotted in
Fig.~\ref{fig:finalplot}, together with the measured widths. In the same
Figure we also show the result of the calculation if we do not take the
level shifts into account. Clearly, the effect of the level shifts on
the observed linewidth is comparable with the direct broadening effect.

From the Figure we also see that the observed broadening of up to about
25\% is larger than the calculated broadening by about a factor two.
This cannot be explained by the most obvious sources of spurious
broadening. These include Doppler broadening ($<2$\%), Zeeman broadening
due to a spurious magnetic field ($<3$\%), and power broadening
($<0.5$\%). Furthermore these broadening mechanisms do not show the
observed signature of increasing with the angle of incidence. A drift of
the laser frequency can be excluded by the same argument, plus the
free-space data point.

A possible mechanism that would have the correct signature is transit
time broadening. To investigate this we numerically integrated the
time-dependent optical Bloch equations for an atom moving through the EW
field. We made the approximation that the atom is a two-level system.
First, the known ground state level shift was used to solve for the
accelerated motion $z(t)$ towards the surface. This solution was then
used to define a time-dependent Rabi-frequency
$\Omega_1(z(t))=\Omega_{10}\exp(-2z(t)/\xi)$, and similarly for the
detuning $\delta(z(t))$, and radiative linewidth $\Gamma(z(t))$. Using
these time dependent parameters we numerically integrated the optical
Bloch equations to obtain the time evolution of the Bloch vector
$(u(t),v(t),w(t))$. Note that power broadening is naturally included in
this method. The number of photons scattered by the atom on its way down
to the surface was obtained as $\int\Omega_1(t)v(t)dt$
\cite{B:CohDupGry92}. Again an averaging over the magnetic $m$-levels
was performed. Finally the probe detuning was varied and a Lorentzian
fit to the obtained absorption profile was performed, as before.
The results of the Bloch equation approach are also shown in
Fig.~\ref{fig:finalplot}. The two calculations yield very similar
results. This shows that transit time broadening does not explain the
discrepancy between calculations and measurements.

As a tentative explanation we invoke the presence of local Stark shifts
caused by charged or polarized particles on the surface. Based on a
straightforward model calculation we find that a surface charge density
of $45e/\lambda^2$ yields a 10\% linewidth increase. Remarkably, such a
charge density corresponds to an average distance between the charges of
order $\sim 100\,$nm, which is just the distance scale to which our
experiment is very sensitive. These calculations only weakly reproduce
the angular dependence shown by the data. Recently McGuirk {\em et al.\
} have reported that Rb adsorbed on a Si or Ti surface generates local
Stark shifts that were measurable as a change in the trapping frequency
of their magnetic trap \cite{McGHarObr04}. The authors mention that
similar effects on a glass surface like ours are very small. However,
their experiment measured only {\em changes} upon depositing clouds of
Rb atoms on the surface, whereas our experiment is also sensitive to
statically present adsorbates. Furthermore, there may be other charged
or polarized adsorbates on the surface. For these reasons local Stark
shifts do seem to present a likely mechanism to explain our results. Our
experiment is also complementary to Ref.~\cite{McGHarObr04} in the sense
that the latter measures a global effect, whereas our experiment is
sensitive only to local variations of the electric fields. Unfortunately
we have no detailed information about possible adsorbates to make a more
quantitative analysis.

\section{Conclusion}

In conclusion, we have observed a broadening of the absorption linewidth
of the $D_2$ resonance line of $^{87}$Rb, caused by the surface. Part of
the broadening can be explained as a combined effect of CQED linewidth
broadening and level shifts due to the proximity of a dielectric
surface. The observed broadening of up to 25\% was about twice that
expected from CQED calculations. The likely candidate to explain this
discrepancy are local Stark shifts due to charged or polarized
adsorbates on the surface.

\begin{acknowledgments}

This work is part of the research program
of the Stichting voor Fundamenteel Onderzoek van de Materie (Foundation
for the Fundamental Research on Matter) and was made possible by
financial support from the Nederlandse Organisatie voor Wetenschappelijk
Onderzoek (Netherlands Organization for the Advancement of Research).

\end{acknowledgments}


\end{document}